\documentclass[aps,prl,reprint,groupedaddress]{revtex4-1}
\pdfoutput=1 
\usepackage{graphicx}
\usepackage{color}
\newcommand{\beq}{\begin{equation}}
\newcommand{\eeq}{\end{equation}}
\newcommand{\bea}{\begin{eqnarray}}
\newcommand{\eea}{\end{eqnarray}}
\newcommand \hmu {\hat{\mu}}
\begin{document}
\title{Kaon fluctuations from lattice QCD}
% repeat the \author .. \affiliation  etc. as needed
% \email, \thanks, \homepage, \altaffiliation all apply to the current
% author. Explanatory text should go in the []'s, actual e-mail
% address or url should go in the {}'s for \email and \homepage.
% Please use the appropriate macro foreach each type of information

% \affiliation command applies to all authors since the last
% \affiliation command. The \affiliation command should follow the
% other information
% \affiliation can be followed by \email, \homepage, \thanks as well.
\author{J. Noronha-Hostler$^a$, R. Bellwied$^a$, J. G\"unther$^b$, P. Parotto$^a$, A. Pasztor$^b$, I. Portillo Vazquez$^a$, C. Ratti$^a$}
%\homepage[]{Your web page}
%\thanks{}
%\altaffiliation{}
\affiliation{\small{\it $^a$ Department of Physics, University of Houston, Houston, TX 77204, USA}}
\affiliation{\small{\it $^b$ Department of Physics, University of Wuppertal, Gaussstr. 20, D-42119 Wuppertal, Germany}}
%Collaboration name if desired (requires use of superscriptaddress
%option in \documentclass). \noaffiliation is required (may also be
%used with the \author command).
%\collaboration can be followed by \email, \homepage, \thanks as well.
%\collaboration{}
%\noaffiliation

\date{\today}
\
\begin{abstract}
We show that it is possible to isolate a set of kaon fluctuations in lattice QCD. By means of the Hadron Resonance Gas (HRG) model, we calculate the actual kaon second-to-first fluctuation ratio, which receives contribution from primordial kaons and resonance decays, and show that it is very close to the one obtained for primordial kaons in the Boltzmann approximation. The latter only involves the strangeness and electric charge chemical potentials, which are functions of $T$ and $\mu_B$ due to the experimental constraint on strangeness and electric charge, and can therefore be calculated on the lattice. This provides an unambiguous method to extract the kaon freeze-out temperature, by comparing the lattice results to the experimental values for the corresponding fluctuations.
% insert abstract here
\end{abstract}
% insert suggested PACS numbers in braces on next line
% insert suggested keywords - APS authors don't need to do this
%\keywords{}

%\maketitle must follow title, authors, abstract, \pacs, and \keywords
\maketitle
Heavy ion collision experiments at RHIC and the LHC recreate in the lab the Quark-Gluon Plasma (QGP), the deconfined phase of strongly interacting matter which exists under extreme conditions of temperatures or densities. While the LHC is focused on the low-density, high-temperature region, the finite density part of the QCD phase diagram is being explored experimentally at RHIC by means of the Beam Energy Scan. Indeed, it is possible to increase the net baryonic density created in an event by decreasing the collision energy, thus, reaching unexplored regions in the $(T, \mu_B)$ plane. From the theoretical point of view, several approaches are being developed to supplement or interpret the experimental information. Lattice QCD is the most reliable first principles method to solve the fundamental theory of strong interactions in its non-perturbative regime. Nevertheless, actual simulations at finite density are not possible at the moment, due to the sign problem. It is, however, possible to reach small values of the chemical potentials thanks to different approaches, such as analytical continuation from imaginary chemical potentials \cite{deForcrand:2002hgr,D'Elia:2002gd,Wu:2006su,D'Elia:2007ke,Conradi:2007be,deForcrand:2008vr,D'Elia:2009tm,Moscicki:2009id}, multiparameter reweighting techniques \cite{Fodor:2001au,Fodor:2001pe,Csikor:2004ik,Fodor:2004nz} and the Taylor expansion of the thermodynamic observables around $\mu_B=0$ \cite{Allton:2002zi,Allton:2005gk,Gavai:2008zr,Basak:2009uv,Kaczmarek:2011zz}. Thermodynamic quantities are therefore becoming available also at finite $\mu_B$.

The observable of conserved charges fluctuations has recieved much interest in recent years. The reason for this is that they can be simulated on the lattice as well as be measured in heavy ion collision experiments. A comparison between theoretical simulations and experimental results allows for the extraction of information about the system created in the lab from first principles. In particular, higher order fluctuations are very sensitive to the critical point such that they have long been understood as a fundamental observable to locate its position\ \cite{Stephanov:1999zu,Gavai:2008zr,Cheng:2007jq}. Lower order fluctuations can be measured and simulated with great accuracy and allow a precise determination of the chemical freeze-out temperature and chemical potential \cite{Karsch:2012wm,Bazavov:2012vg,Borsanyi:2013hza,Borsanyi:2014ewa,Bazavov:2015zja}.  By analyzing the baryon number and electric charge fluctuations separately, a consistent scenario emerges in Ref. \cite{Borsanyi:2014ewa}, which gives rise to the same freeze-out temperature and chemical potential for these two conserved charges. However, strangeness is still missing from the picture mainly due to the lack of experimental data for the fluctuations of (multi-)strange baryons. Therefore, it is important to have a first-principle determination of the strangeness freeze-out temperature. The ALICE data for particle yields and ratios seem to indicate a tension between the freeze-out temperatures in the light and strange sectors \cite{Preghenella:2011np,Floris:2014pta}. Several explanations have been proposed for this result \cite{Becattini:2012sq,Becattini:2012xb,Steinheimer:2012rd,Bellwied:2013cta,Bazavov:2014xya,Noronha-Hostler:2014usa,Noronha-Hostler:2014aia,Chatterjee:2013yga}, but so far none has been validated or excluded. A result from first principles would finally resolve this issue.

Preliminary results of kaon fluctuations have been presented by the STAR collaboration at the Quark Matter 2012 \cite{McDonald:2012ts}, Quark Matter 2015 \cite{Thader:2016gpa} and Strangeness in Quark Matter 2016 \cite{SQM} conferences, which will soon be finalized. Since kaons carry strangeness, extracting their freeze-out temperature from first principles constitutes a first important step towards understanding the freeze-out of  strangeness in heavy ion collisions.
The experimental kaon distribution includes contributions from both primordial kaons and the feed down of resonance decays, and it is subjected to rapidity and transverse momentum cuts. The resonance decays in principle cannot be captured by a thermal calculation in equilibrium, such as lattice QCD. Since resonance decays are governed by branching ratios which appear as factors in front of the corresponding conserved charge in the fluctuations of the daughter particles, the latter can be very different when they are calculated in equilibrium or taking the resonance decays into account. Due to charge conservation, fluctuations of a conserved charge would be the same in a fully equilibrated system and in a system which allows for resonance decays, but this is not the case for the fluctuations of a specific particle. To quantify this effect for kaons, we use the Hadron Resonance Gas (HRG) model to show that  the resonance decays play no role for the lower order fluctuations. Also, the fluctuations of primordial kaons in the Boltzmann approximation coincide with the ones of the full distribution, including resonance decays. %We address the issue of the rapidity and transverse momentum cuts and check their effects on the fluctuations. 
We repeat the same analysis also for the protons: we check the validity of the Boltzmann vs. the full distribution including resonance decays as well as isospin randomization.

The HRG model is based on the idea that a gas of interacting hadrons in their ground state can be well described by a gas of non-interacting hadrons and resonances \cite{Dashen:1969ep,Venugopalan:1992hy,Karsch:2003vd,Karsch:2003zq,Tawfik:2004sw,Ratti:2010kj}. The partition function of the model, therefore, can be written as a sum of ideal gas contributions of all known baryons and mesons:
\bea
\frac{p^{HRG}}{T^4}&=&\frac{1}{VT^3}\sum_{i\in mesons}\ln Z^M_{i}(T,\mu_{X^a})
\nonumber\\
&+&\frac{1}{VT^3}\sum_{i\in baryons}\ln Z^{B}_{i}(T,\mu_{X^a})
\eea
where
\bea
&&\ln Z^{M/B}_{i}=\mp \frac{V d_i}{2\pi^2}\int_0^{\infty}dk k^2\ln(1\mp z_i\exp{-\epsilon_i/T})
\\
&&\epsilon_i=\sqrt{k^2+m_i^2},~~~~~~~~~~~~z_i=\exp((\sum_a X_i^a\mu_{X^a})/T)
\nonumber
\eea
and $X^a$ are the conserved charges, namely baryon number $B$, electric charge $Q$ and strangeness $S$.
For each particle, it is possible to write the partition function as:
\bea
\ln Z_i^{M/B}&\simeq&\frac{d_i}{2\pi^2}\left(\frac{m_i}{T}\right)^2\sum_{k=1}^{\infty}\frac{(\pm1)^{k+1}}{k^2}K_2\left(\frac{km_i}{T}\right)	\times
\nonumber\\
&\times&\cosh\left[k\left(B_i\mu_B+Q_i\mu_Q+S_i\mu_S\right)/T\right].
\label{Boltzmann}
\eea
In Eq. (\ref{Boltzmann}), $k=1$ corresponds to the Boltzmann approximation. In this limit, it is possible to isolate the contributions of different hadrons to the total pressure, distinguishing them through their quantum numbers. This was done in Ref. \cite{Bazavov:2013dta} in order to distinguish the contributions to the pressure of strange mesons and (multi-) strange baryons:
\begin{eqnarray}
P_{S}(\hmu_B,\hmu_S) &=& P_{0|1|} \cosh(\hmu_S) 
\nonumber \\
&+& P_{1|1|} \cosh(\hmu_B-\hmu_S)
\nonumber \\
&+& P_{1|2|} \cosh(\hmu_B-2\hmu_S)
\nonumber \\
&+& P_{1|3|} \cosh(\hmu_B-3\hmu_S) 
\;,
\end{eqnarray}
where dimensionless chemical potentials are used $\hmu_{B/S}=\mu_{B/S}/T$.  Here $P_{0|1|}$ is the partial pressure of all strange
$|S|=1$ mesons and for the baryons $P_{1,|i|}$ are the partial pressures of all $|S|=i$
($i=1,2,3$) baryons.  Each of the $P_{B|S|}$ coefficents are defined via susceptibilities:
\begin{eqnarray}
P_{0|1|} &=& \chi_2^S -\chi_{22}^{BS} 
\;, \label{eq:M} \\
P_{1|1|} &=&  \frac{1}{2} \left( \chi_4^S - \chi_2^S +5 \chi_{13}^{BS}+
7 \chi_{22}^{BS} \right) 
\;, \label{eq:B1} \\
P_{1|2|} &=& - \frac{1}{4} \left( \chi_4^S - \chi_2^S + 4 \chi_{13}^{BS} +
4 \chi_{22}^{BS} \right)
\;, \label{eq:B2}\\
P_{1|3|} &=& \frac{1}{18} \left( \chi_4^S -  \chi_2^S + 3 \chi_{13}^{BS}+
3 \chi_{22}^{BS} \right) \;, \label{eq:B3}
\end{eqnarray}
where we set $c_1=0$ and $c_2=0$ from the original paper \cite{Bazavov:2013dta} since we are only interested in the hadronic sector of the equation of state. The susceptibilities are defined as
\begin{equation}
\chi_{mnk}^{BSQ} = \left. \frac{\partial^{(m+n+k)} [p(\hmu_B,\hmu_S,\hmu_Q)/T^4]} 
{\partial \hmu_B^m \partial \hmu_S^n \hmu_Q^k} \right|_{\vec{\mu}=0}
\ .
\label{eq:susc}
\end{equation}

Derivatives of the pressure with respect to the strange chemical potential then follow via
\begin{equation}
\chi^S_n (M,|S|=1)=\left(\chi_2^S -\chi_{22}^{BS}\right)|_{\vec{\mu}=0} \frac{ d^n\cosh(\hmu_S)}{d \hmu_S^n},
\end{equation}
which implies that the prefactor cancels when ratios of the derivatives are taken.

Note that  in \cite{Bazavov:2013dta} the effect of electric charge was not considered. However, here we find that the conservation of electric charge is necessary if one wants to consider a large baryon chemical potential. To do so, one must rewrite the full pressure separating by baryon number, strangeness, and charge in the following way
%\begin{widetext}
\begin{eqnarray}\label{eqn:Pbsq}
P(\hmu_B,\hmu_S,\hmu_Q) &=& P_{000}+ P_{00|1|}\cosh(\hmu_Q) 
+ P_{100}\cosh(\hmu_B)
\nonumber\\&+&P_{101}\cosh(\hmu_B+\hmu_Q)+P_{10-1}\cosh(\hmu_B-\hmu_Q)\nonumber\\
&+&P_{102}\cosh(\hmu_B+2\hmu_Q)+P_{0|1|0}\cosh(\hmu_S)
\nonumber\\&+&P_{0|1||1|}\cosh(\hmu_S+\hmu_Q)+P_{1|1|0} \cosh(\hmu_B-\hmu_S)\nonumber\\
&+&P_{1|1|1} \cosh(\hmu_B-\hmu_S+\hmu_Q)
\nonumber\\&+&P_{1|1|-1} \cosh(\hmu_B-\hmu_S-\hmu_Q)
\nonumber\\&+&P_{1|2|0} \cosh(\hmu_B-2\hmu_S)\nonumber\\
&+&P_{1|2||1|} \cosh(\hmu_B-2\hmu_S-\hmu_Q)
\nonumber\\&+&P_{1|3||1|} \cosh(\hmu_B-3\hmu_S-\hmu_Q).
\end{eqnarray}
%\end{widetext}

In the above formula, we can identify the contribution of charged kaons and their resonances to the pressure as:
\bea
P_{K^{+/-}}=P_{0|1||1|}\cosh(\hat{\mu}_S+\hat{\mu}_Q)
\eea

While the form of Eq.\ (\ref{eqn:Pbsq}) is quite complicated, the derivatives of the partial pressures then follow such that
\begin{eqnarray}
\chi^S_e(BS)&=&\chi^B_e(BS)=\chi^{S/B}_2(BS)=\chi^{S/B}_4(BS)\ldots \nonumber\\
\chi^B_o(BS)&=&\chi^B_1(BS)=\chi^B_3(BS)\ldots \nonumber\\ 
\chi^S_o(BS)&=&\chi^S_1(BS)=\chi^S_3(BS)\ldots \nonumber\\ 
\label{eqn:supn}
\end{eqnarray}
where $S/B$ indicates either the derivative respective to strangeness or baryon number, respectively, and $e$ or $o$ indicate even or odd derivatives.  
Taking the ratios of $\chi^S_n/\chi^S_m$ the prefactor of susceptibilities shown in Eqs. (\ref{eq:M}-\ref{eq:B3}) cancel out, for example for net-kaons:
\begin{eqnarray}
\label{fluc_S}
\frac{\chi^K_e}{\chi^K_o}&=&\frac{\cosh(\hmu_S+\hmu_Q)}{\sinh(\hmu_S+\hmu_Q)}\nonumber\\
\frac{\chi^K_o}{\chi^K_e}&=&\frac{\sinh(\hmu_S+\hmu_Q)}{\cosh(\hmu_S+\hmu_Q)}\nonumber\\
\frac{\chi^K_o}{\chi^K_o}&=&\frac{\chi^K_e}{\chi^K_e}=1
\end{eqnarray}
Similarly, for net-protons one finds

\begin{eqnarray}\label{fluc_p}
\frac{\chi^p_e}{\chi^p_o}&=&\frac{\cosh(\hmu_B+\hmu_Q)}{\sinh(\hmu_B+\hmu_Q)}
\nonumber\\
\frac{\chi^p_o}{\chi^p_e}&=&\frac{\sinh(\hmu_B+\hmu_Q)}{\cosh(\hmu_B+\hmu_Q)}\nonumber\\
\frac{\chi^p_o}{\chi^p_o}&=&\frac{\chi^p_e}{\chi^p_e}=1.
\end{eqnarray}

In the following, we will use the above formulas for fluctuations of particles containing a given set of quantum number (e.g. kaons) and compare them to the actual particle fluctuations, which take into account the contribution of primordial distributions and resonance decays.
 Taking the effect of decays into account as shown in  \cite{Alba:2014eba,Alba:2015iva}, an extra term is included in front of the susceptibility such that:
\begin{widetext}
\begin{equation}
\tilde{\chi}_n^{S}=\sum_j^{X_{stable}}\sum_i^{N_{HRG}}\left(Pr_{ij} S_j\right)^n \frac{g_i}{2\pi^2}\frac{\partial^n}{\partial \mu_S^n}\left\{\int_0^{\infty} dp_T \frac{p_T^2}{\left(Exp\left[\sqrt{p_T^2+m^2_i}-(B_i\mu_b+S_i\mu_S+Q_i\mu_Q)\right]+(-1)^{B_i+1}\right)}\right\}\label{eqn:chis}
\end{equation}
\end{widetext}
where $Pr_{ij}$ is the probility for a resonance $i$ to produce a daughter particle $j$.  Note that this is slightly different than a branching ratio.  A branching ratio is the probability of a resonance $i$ decaying into a specific decay channel.  Here $Pr_{ij}$ includes all decays channels for a resonance such that
\begin{equation}
Pr_{ij}=\sum_{c}Br_{i\rightarrow c} n_j(c)
\end{equation}
where $Br_{i\rightarrow c}$ is the branching for the resonance $i$ to decay into the stable decay product $c$ and $n_j(c)$ is the number of times particle $j$ appears in channel $c$.  For instance, if we observe the decay channel $i\rightarrow K+\pi^+ + \pi^+$, then $n_{\pi^+}=2$ whereas for the same decay channel $n_{\pi^-}=0$. Note that this includes all subsequent decays until only the stable particles remain.

In Eq. (\ref{eqn:chis}), $X_{stable}$ is the sum over the stable particles that one is observing.  For the case of net-strangeness one would typically consider $K^+$ and $K^-$, however, it may be possible to eventually include $K^0$ and $\bar{K}^0$ experimentally.  In this paper we will consider the contributions of only  $K^+$ and $K^-$.

\begin{figure}[h]
\hspace{-.5cm}
\begin{minipage}{0.24\textwidth}
 \scalebox{.34}{
 \includegraphics{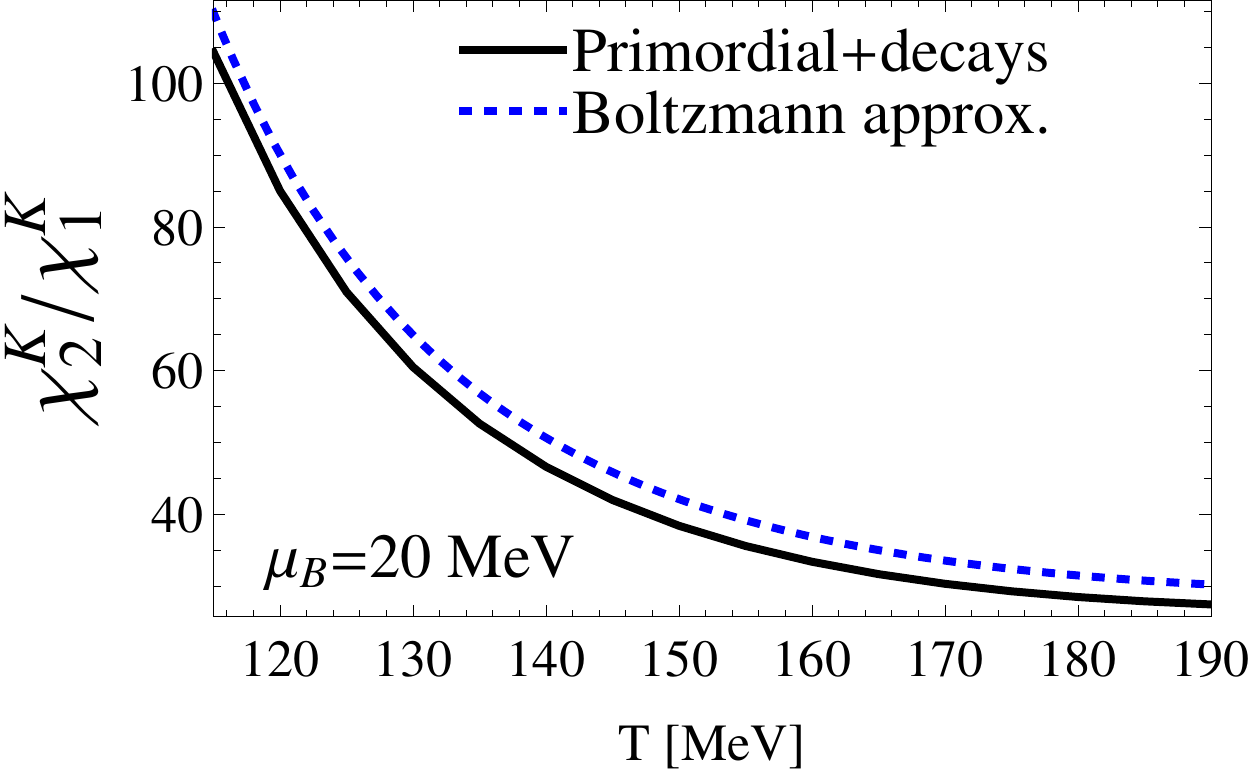}}
 \end{minipage}
%\hspace{-2cm}
\begin{minipage}{0.24\textwidth}
 \scalebox{.34}{
 \includegraphics{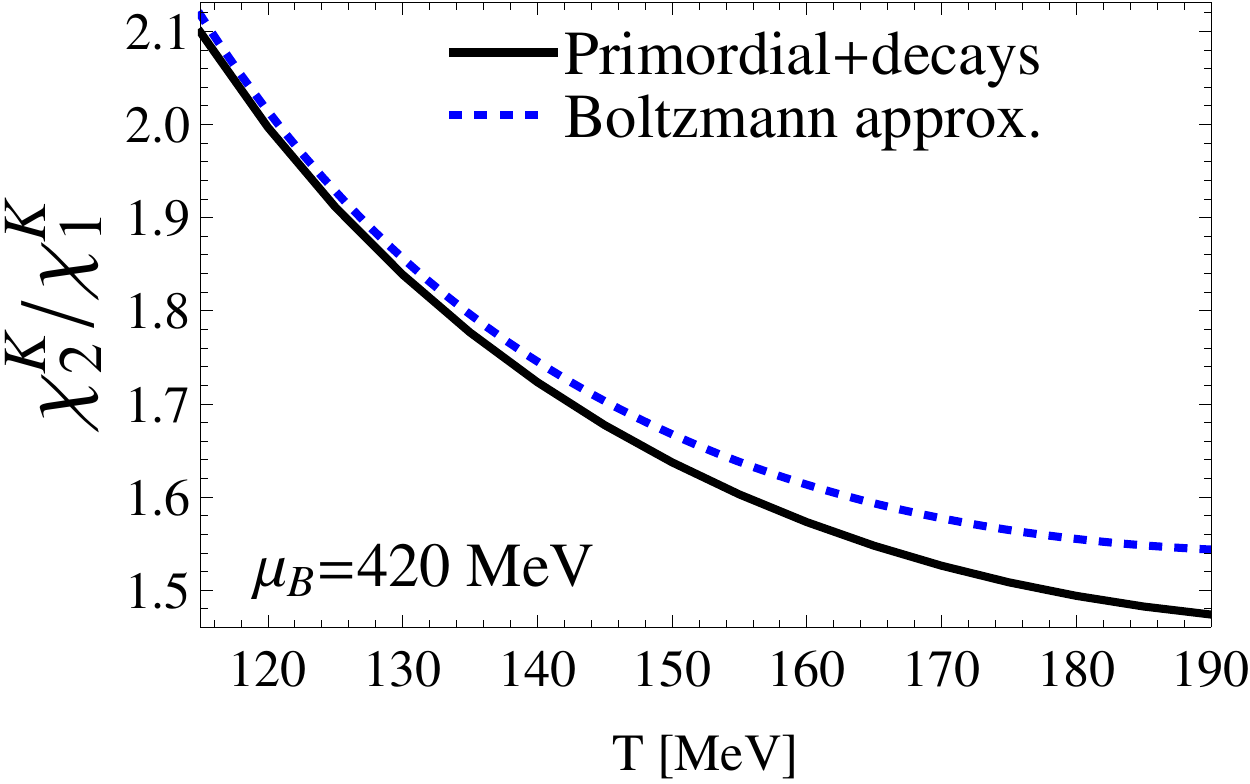}}
\end{minipage}\\
\hspace{-.5cm}
\begin{minipage}{0.24\textwidth}
 \scalebox{.34}{
 \includegraphics{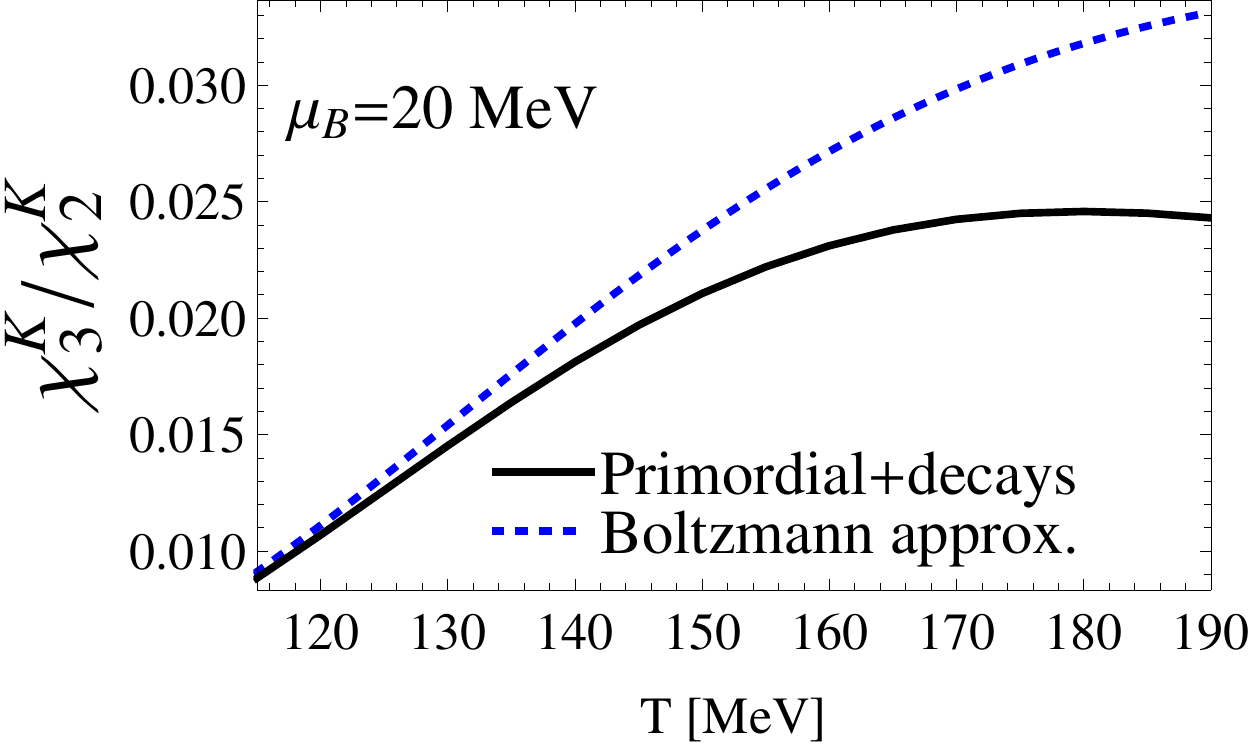}}
 \end{minipage}
%\hspace{-2cm}
\begin{minipage}{0.24\textwidth}
 \scalebox{.34}{
 \includegraphics{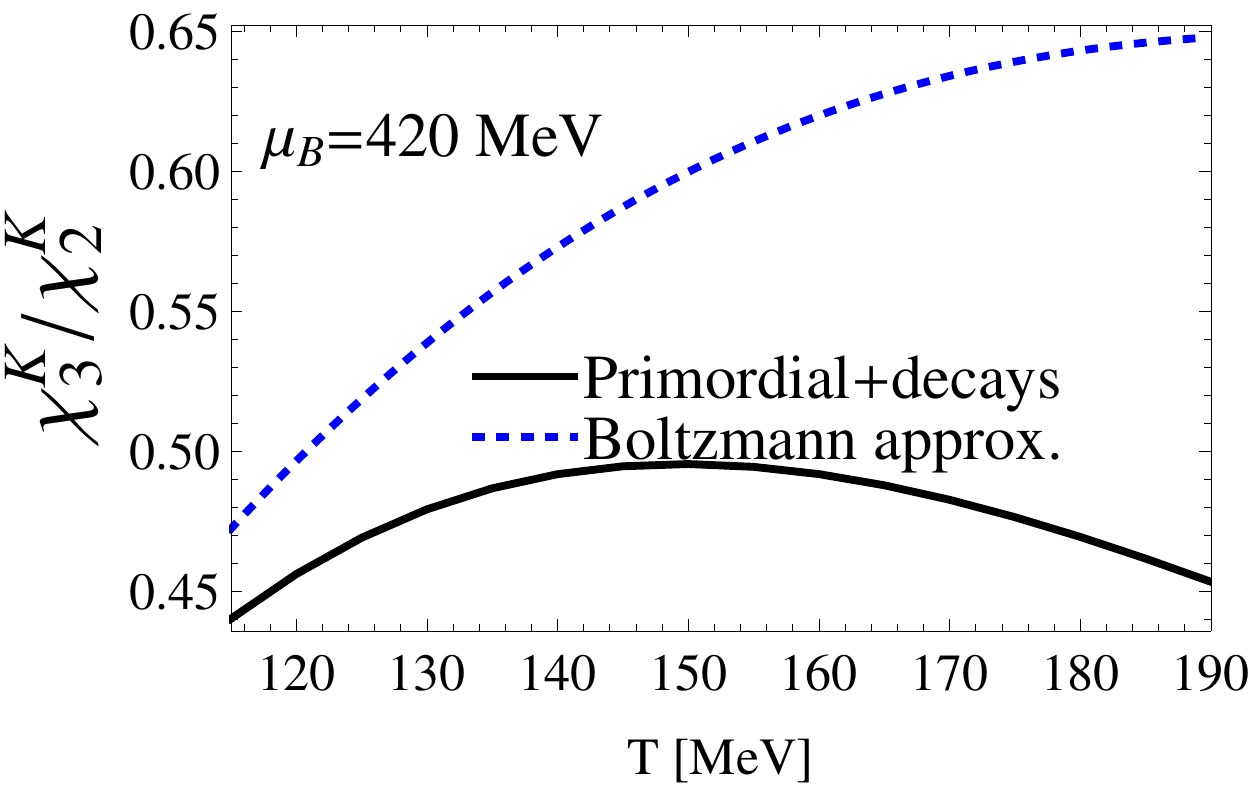}}
\end{minipage}
\caption{Upper panels: comparison between $\chi_2^K/\chi_1^K$ for primordial kaons + resonance decays (full, black line), to the ones obtained from the upper Eq. (\ref{fluc_S}) (dashed, blue line). The two figures correspond to two different chemical potentials: $\mu_B=20$ MeV (left) and $\mu_B=420$ MeV (right). Lower panels: comparison between $\chi_3^K/\chi_2^K$ for primordial kaons + resonance decays (full, black line), to the ones obtained from the middle Eq. (\ref{fluc_S}) (dashed, blue line). The two figures correspond to two different chemical potentials: $\mu_B=20$ MeV (left) and $\mu_B=420$ MeV (right). \label{fig1}}
\end{figure}

\begin{figure}
\hspace{-.5cm}
\begin{minipage}{0.48\textwidth}
 \scalebox{.7}{
 \includegraphics{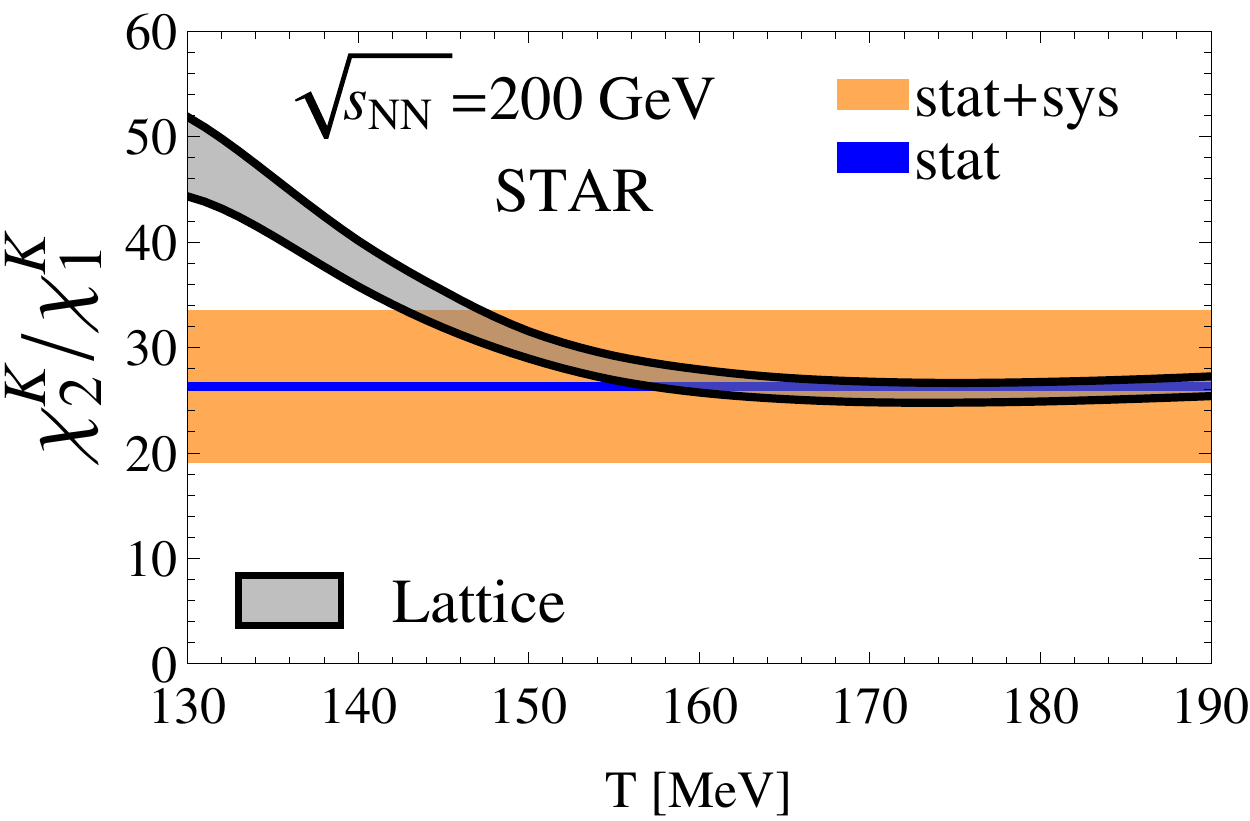}}
 \end{minipage}
\caption{Example of comparison between $\chi_2^K/\chi_1^K$ from lattice QCD, calculated from the upper Eq. (\ref{fluc_S}), and the preliminary STAR result at $\sqrt{s}=200$ GeV shown at the Strangeness in Quark Matter 2016 conference \cite{SQM}. The blue band corresponds to the statistical error, the orange one shows statistical and systematic errors summed in quadrature. \label{fig2}}
\end{figure}

In Fig. \ref{fig1} we show $\chi_2^K/\chi_1^K$ (upper panels) and $\chi_3^K/\chi_2^K$ (lower panels) for kaons. The black, solid curves are the actual kaon fluctuations obtained in the HRG model including primordial kaons and the feed down from resonance decays. The blue, dashed curves correspond to the upper Eq. (\ref{fluc_S}). The left panels are calculated at $\mu_B=20$ MeV, the right ones at $\mu_B=420$ MeV. It is evident that, for the lower fluctuation ratios, the Boltzmann approximation yields a very good description of the actual curve, both for low and high chemical potentials, which cover the entire range spanned by the RHIC Beam Energy Scan. This means that it is safe to use the upper Eq. (\ref{fluc_S}) to extract the kaon fluctuation ratio $\chi_2^K/\chi_1^K$ from lattice QCD, and compare it to experimental data from RHIC in order to extract the kaon freeze-out temperature. On the other hand, $\chi_3/\chi_2$ shows a discrepancy between the two approaches already for small chemical potentials: this means that the effect of resonance decays, in particular of multi-strange baryons decaying into kaons, significantly affects the higher order fluctuations and their ratios.
In Fig. \ref{fig2} we show an example of comparing the lattice QCD $\chi_2^K/\chi_1^K$ for kaons, obtained from the upper Eq. (\ref{fluc_S}), to the preliminary experimental data from STAR. The accuracy of the lattice QCD data will allow a precise determination of the kaon freeze-out temperature once the experimental error-bars are under control.
%\begin{figure}
%\hspace{-.5cm}
%\begin{minipage}{0.48\textwidth}
% \scalebox{.7}{
% \includegraphics{finalfigs2/accept}}
% \end{minipage}
%\caption{Comparison between $\chi_2^K/\chi_1^K$ with (full, black line) and without (blue, dashed line) experimental cuts at $\mu_B=24.3$ MeV. \label{fig4}}
%\end{figure}
%In Fig. \ref{fig4} we study the effect of including the experimental cuts in rapidity and momentum on the $\chi_2^K/\chi_1^K$, an effect which could affect the experimental distribution but cannot be captured by lattice QCD calculations. While $\chi_1^K$ and $\chi_2^K$ are separately affected by the cuts, these effects cancel out in the ratio such that the two curves in Fig. \ref{fig4}, with and without cuts, are essentially on top of each other.
\begin{figure}
\hspace{-.5cm}
\begin{minipage}{0.24\textwidth}
 \scalebox{.34}{
 \includegraphics{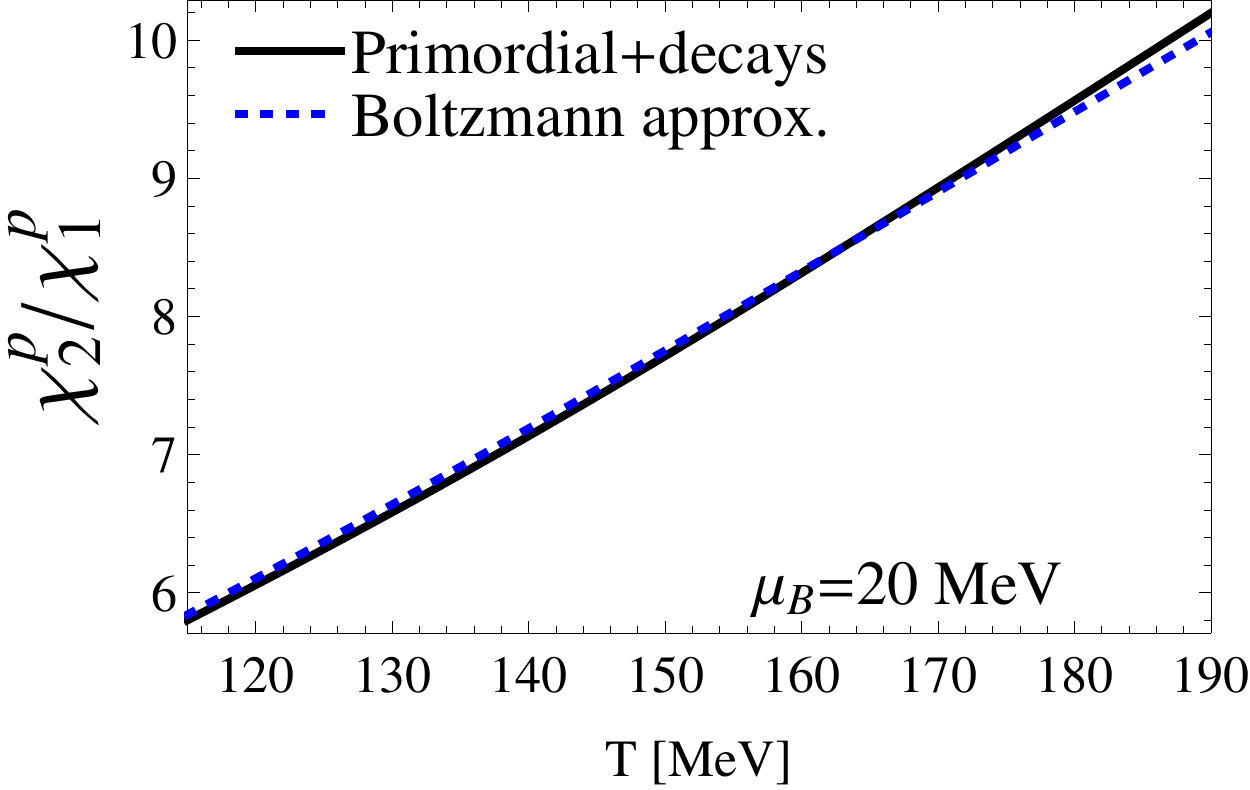}}
 \end{minipage}
%\hspace{-2cm}
\begin{minipage}{0.24\textwidth}
 \scalebox{.34}{
 \includegraphics{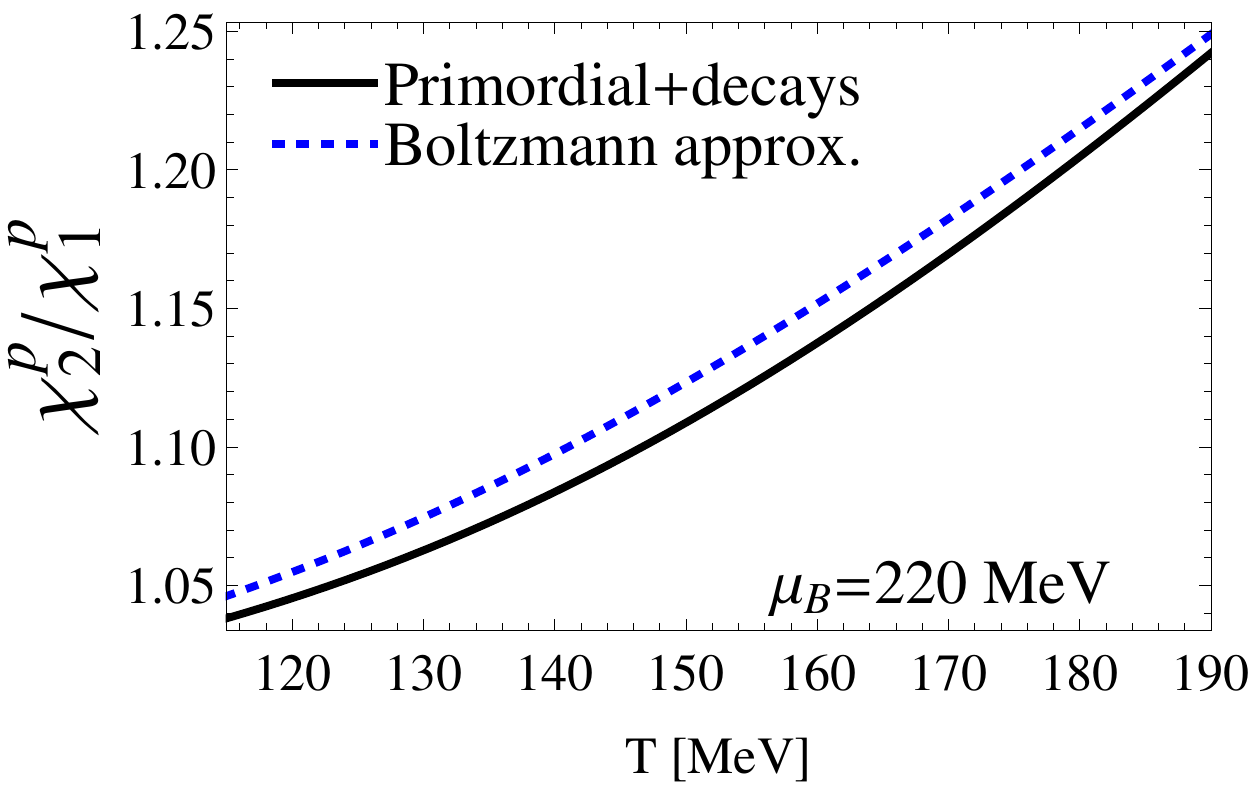}}
\end{minipage}\\
\hspace{-.5cm}
\begin{minipage}{0.24\textwidth}
 \scalebox{.34}{
 \includegraphics{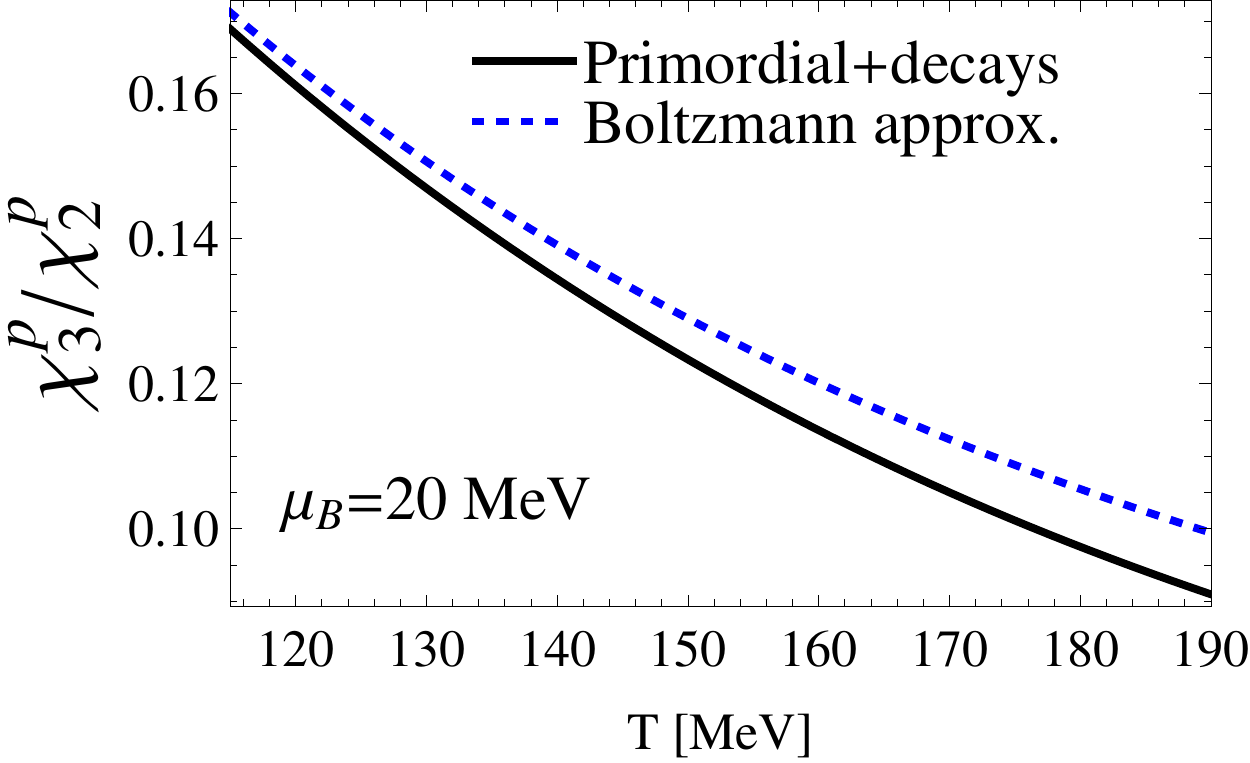}}
 \end{minipage}
%\hspace{-2cm}
\begin{minipage}{0.24\textwidth}
 \scalebox{.34}{
 \includegraphics{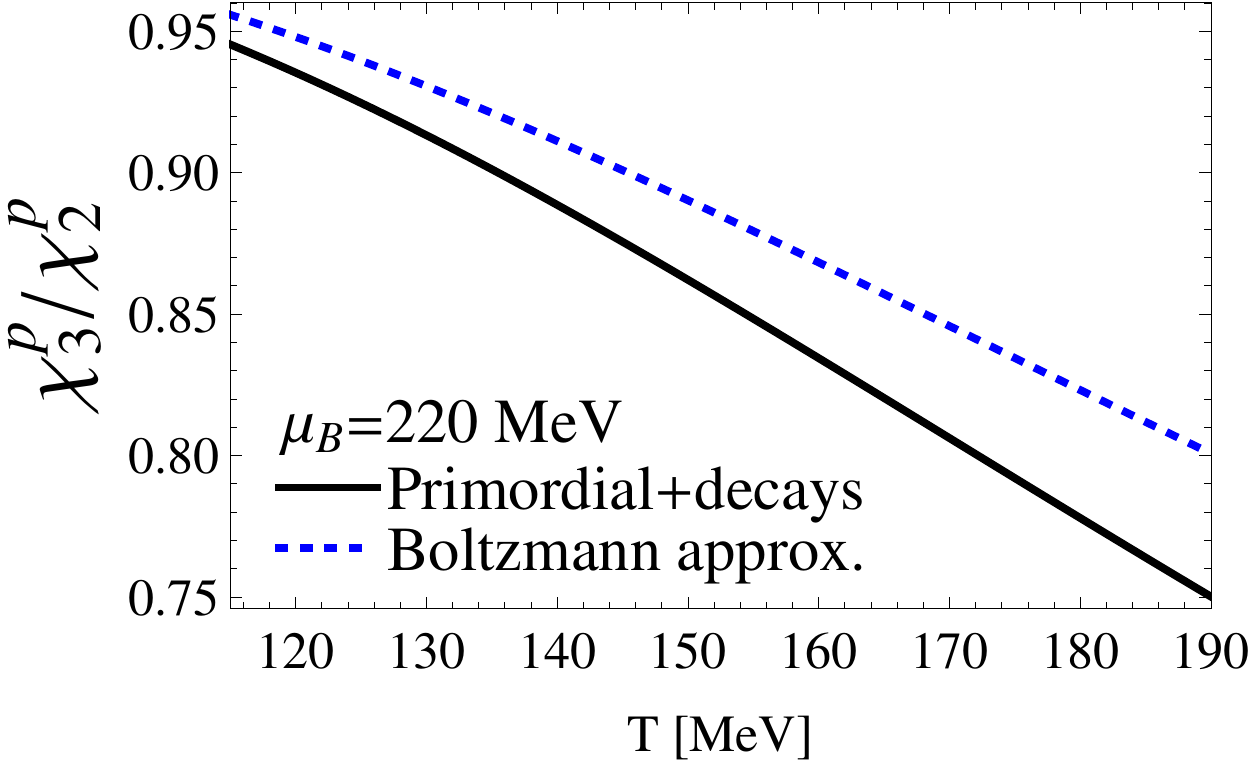}}
\end{minipage}
\caption{Upper panels: comparison between $\chi_2^p/\chi_1^p$ for primordial protons + resonance decays (full, black line), to the ones obtained from the upper Eq. (\ref{fluc_p}) (dashed, blue line). The two figures correspond to two different chemical potentials: $\mu_B=20$ MeV (left) and $\mu_B=220$ MeV (right). Lower panels: comparison between $\chi_3/^p\chi_2^p$ for primordial protons + resonance decays (full, black line), to the ones obtained from the middle Eq. (\ref{fluc_p}) (dashed, blue line). The two figures correspond to two different chemical potentials: $\mu_B=20$ MeV (left) and $\mu_B=220$ MeV (right). \label{fig3}}
\end{figure}

In Fig. \ref{fig3} we show $\chi_2^p/\chi_1^p$ (upper panels) and $\chi_3^p/\chi_2^p$ (lower panels) for protons. The black, solid curves are the actual proton fluctuations, obtained in the HRG model including primordial protons and those coming from resonance decays with isospin randomization, as described in Ref. \cite{Nahrgang:2014fza}. The blue, dashed curves correspond to the upper Eq. (\ref{fluc_p}). The left panels are calculated at $\mu_B=20$ MeV, the right ones at $\mu_B=220$ MeV. It is evident that, for the lower fluctuation ratios, the Boltzmann approximation yields a good description of the actual curve, both for low and high chemical potentials. Deviations start to occur at $\mu_B\simeq300$ MeV, which allows a comparison with the experimental data for $\sqrt{s}\leq14.5$ GeV \cite{Alba:2014eba}. On the other hand, $\chi_3/\chi_2$ shows a $\sim$15\% discrepancy between the two approaches in the temperature range of interest.

In conclusion, using the Hadron Resonance Gas model we have shown that it is safe to use the Boltzmann approximation to extract $\chi_2/\chi_1$ for kaons (and protons) from lattice QCD. The results presented here are relying on the assumption that the HRG model is a valid approximation of QCD at the chemical freeze-out, which is reasonable for lower order fluctuations. The curves corresponding to the Boltzmann approximation are very close to the actual ones, which take into account not only the primordial particles but also those produced by resonance decays, for a large range of chemical potentials which cover most of the energies of RHIC BES. This will allow safe extraction of the kaon freeze-out parameters once the experimental data for kaon fluctuations are finalized. The implications of the difference (or lack thereof) between the light and strange freeze-out parameters at the Beam Energy Scan may have far reaching effects within hydrodynamical modeling as well as for the temperature dependence of the baryon and strange diffusion transport coefficients. 
\section*{Acknowledgements}
We would like to thank Volker Koch, Paolo Alba, Szabolcs Borsanyi and Jorge Noronha for fruitful discussions. This material is based upon work supported by the National Science Foundation under grant no. PHY-1513864 and by the U.S. Department of Energy, Office of Science, Office of Nuclear Physics, within the framework of the Beam Energy Scan Theory (BEST) Topical Collaboration. The work of R. Bellwied is supported through DOE grant
DEFG02-07ER41521. This work contains lattice QCD data provided by the Wuppertal-Budapest Collaboration. An  award  of  computer  time  was  provided
by the INCITE program.  This research used resources of the Argonne Leadership Computing Facility, which is a DOE Office of Science User Facility supported under Contract DE-AC02-06CH11357. The work of J. G. and A. P. was supported by the DFG grant SFB/TR55. The authors gratefully acknowledge the Gauss Centre for Supercomputing (GCS) for providing computing time for a GCS Large-Scale Project on the GCS share of the supercomputer JUQUEEN \cite{juqueen} at J\"ulich Supercomputing Centre (JSC). 

\bibliography{Kaon_Fluctuations_v3}
\end{document}